\documentclass[10pt,letterpaper]{article}
\usepackage[top=0.85in,left=2.75in,footskip=0.75in,marginparwidth=2in]{geometry}

\usepackage[utf8]{inputenc}

\usepackage{cite}

\usepackage{nameref,hyperref}

\usepackage[right]{lineno}

\usepackage{microtype}
\DisableLigatures[f]{encoding = *, family = * }

\raggedright
\usepackage{changepage}

\usepackage[aboveskip=1pt,labelfont=bf,labelsep=period,singlelinecheck=off]{caption}

\makeatletter
\renewcommand{\@biblabel}[1]{\quad#1.}
\makeatother

\usepackage{lastpage,fancyhdr,graphicx}
\usepackage{epstopdf}
\fancyheadoffset[L]{2.25in}
\fancyfootoffset[L]{2.25in}

\usepackage{color}

\definecolor{Gray}{gray}{.25}

\usepackage{graphicx}

\usepackage{sidecap}

\usepackage{wrapfig}
\usepackage[pscoord]{eso-pic}
\usepackage[fulladjust]{marginnote}
\reversemarginpar

\begin{document}
\vspace*{0.35in}

\begin{flushleft}
{\Large
\textbf\newline{The possibility for panspermia in the galaxy by means of planetary dust grains}
}
\newline
\\
Zaza N. Osmanov 
\\
\bigskip
School of Physics, Free University of Tbilisi, 0183, Tbilisi, Georgia
\\
E. Kharadze Georgian National Astrophysical Observatory, Abastumani 0301, Georgia
\\
\bigskip
z.osmanov@freeuni.edu.ge

\end{flushleft}

\section*{Abstract}
Under the assumption that planetary dust particles can escape from the gravitational attraction of a planet, we consider the possibility of the dust grains leaving the star’s system by means of radiation pressure. By taking the typical dust parameters into account, we consider their dynamics and show that they can reach the deep space, taking part in panspermia. It has been shown that, during $5$ billion years, the dust grains will reach $10^5$ stellar systems, and by taking the Drake equation into account, it has been shown that the whole galaxy will be full of planetary dust particles. It has been found that dust grains can be trapped within HI and HII clouds of the interstellar medium, suggesting that these regions may harbor not only the building blocks of life but also complex molecules associated with life. We have also emphasized two key factors that hinder the preservation of life within interstellar clouds: (a) the interaction of dust grains with extremely hot regions, which leads to the complete ionization and destruction of complex molecules, and (b) the lack of definitive knowledge regarding the long-term survival of bacteria in the vacuum.


\section{Introduction}

During the last two decades, interest in the search for life and especially in the search for extraterrestrial intelligence (SETI) has increased significantly \cite{shkl}, and several exotic ideas of megastructures of a super-advanced civilization have been proposed \cite{ring,dvali,dsbh}. The possibility of life in deep space has clearly bolstered astrobiological research, becoming one of the major challenges of the modern space sciences \cite{asbio1} \cite{asbio2}.

It is worth noting that the main problem is the origin of life, or abiogenesis, the details of which are still unknown to us. There are two main approaches to this issue: according to one, life originated and developed on Earth, and according to the other approach, which is called panspermia, it originated in deep space and was then transported to Earth via comets or cosmic dust (please see \cite{chandra} and references therein). In this context, it is worth noting that the study of the universality of life has become very popular \cite{schneider}.

The attitude of scientists towards panspermia is not uniform and some believe that it is unlikely, but just possible \cite{acta1}. 
The idea of interplanetary transfer of life has been discussed by Richter in $1865$ and by Lord Kelvin in 1894 \cite{arrhenius}. Richter attempted to combine Darwin's theory of evolution with the conception of panspermia. Lord Kelvin shared a very optimistic view of the theory of panspermia. In his presidential address to the British Association at Edinburgh in $1871$, he says that when a celestial body of more or less the same size as Earth collides with it, fragments containing living organisms will scatter into space and spread life.

The original first scientific review of this problem was considered in \cite{arrhenius}, where the author explored the idea of panspermia only in general terms and gave crude estimates. In particular, it has been pointed out that, by means of solar radiation pressure, small dust grains containing live organisms can travel to the nearest solar system, Alpha Centauri, in nine thousand years. In \cite{shkl} Shklovskii and Sagan have emphasized that for dust grains to travel a thousand light years away from the Earth, hundreds of millions of years will be required. It should be taken into account that during the motion of dust particles in interstellar space, they are under constant ultraviolet radiation. Therefore, bare microorganisms already die very quickly due to exposure to the sun, and for microorganisms "buried" in dust particles, the inactivation time might be of the order of $10^6$ years \cite{UV,wesson}.  Gobart et al. \cite{probab} presented an approach in which the probability of panspermia was investigated. Nonetheless, the probability of panspermia is notably high for only a small fraction of stellar particles.

Berrera has conducted a study on the phenomenon of space dust collisions with atmospheric flow \cite{berera}. The author has shown that a certain fraction of dust grains at high altitudes, after scattering against cosmic dust particles, might be accelerated to escape velocity, leaving the Earth's gravitational field. 

If a similar scenario takes place in other systems, the planetary dust particles, being already free from the planet's gravitational field, might escape from the star's system by means of the radiation pressure and initial velocity, spreading life or organics into the cosmos.

This article is structured as follows: in Section 2, we explore the core concepts and derive the results; Section 3 offers a summary

\begin{figure}
  \centering {\includegraphics[width=8cm]{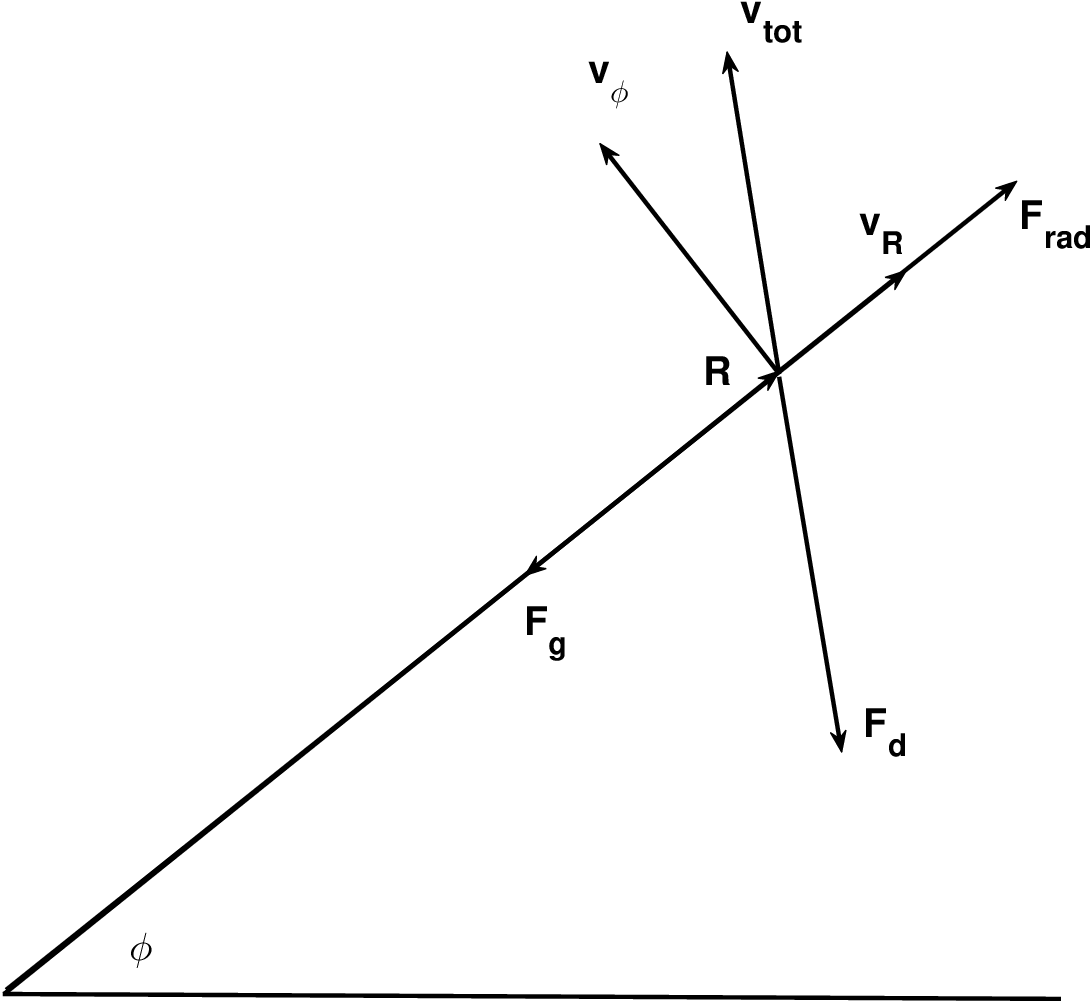}}
  \caption{The schematic picture of the components of the total velocity and forces acting on the particle.}\label{fig1}
\end{figure}

\section{Dynamics of dust particles}

In this section we consider dynamics of dust particles to study how far from a host planet they can travel on a reasonable galactic time scale. The similar problem has been studied by \cite{wesson}, but a task has been performed for a simplified one-dimensional model. In the present manuscript we examine a realistic two-dimensional scenario, by taking into account the rotation of a planet around a host star and a drag force acting on a grain from the interstellar hydrogen medium.

For the dynamics of a planetary dust particle with mass $m$ the equation of motion is given by
\begin{equation}
\label{NL} 
m\frac{d\vec{\upsilon}_{tot}}{dt} = \vec{F_g}+\vec{F}_{rad}+\vec{F_d},
\end{equation}
where $F_g = GMm/R^2$ is the gravitational force, $G$ represents the gravitational constant, $M$ is the star's mass and $R$ is the radial coordinate of the particle (see Fig. 1), $F_{rad}\simeq\frac{Lr^2}{4R^2c}$ denotes the star's radiation force \cite{carroll}, $r$ is the radius of the dust grain (we assume that the dust particles have spherical shapes), $L $ is the star's luminosity, and $c$ represents the speed of light, $F_{d}\approx D\rho\pi r^2\upsilon^2$ denotes the drag force, $D$ is the drag coefficient, $\rho_0\simeq 2m_pn_0$ and $n_0\simeq 1$cm$^{-3}$ are respectively the mass density and the number density of an interstellar medium (ISM), and $m_p$ represents the proton's mass. We assume that the ISM is predominantly composed of molecular hydrogen \cite{shkl}.

\begin{figure}
  \centering {\includegraphics[width=11cm]{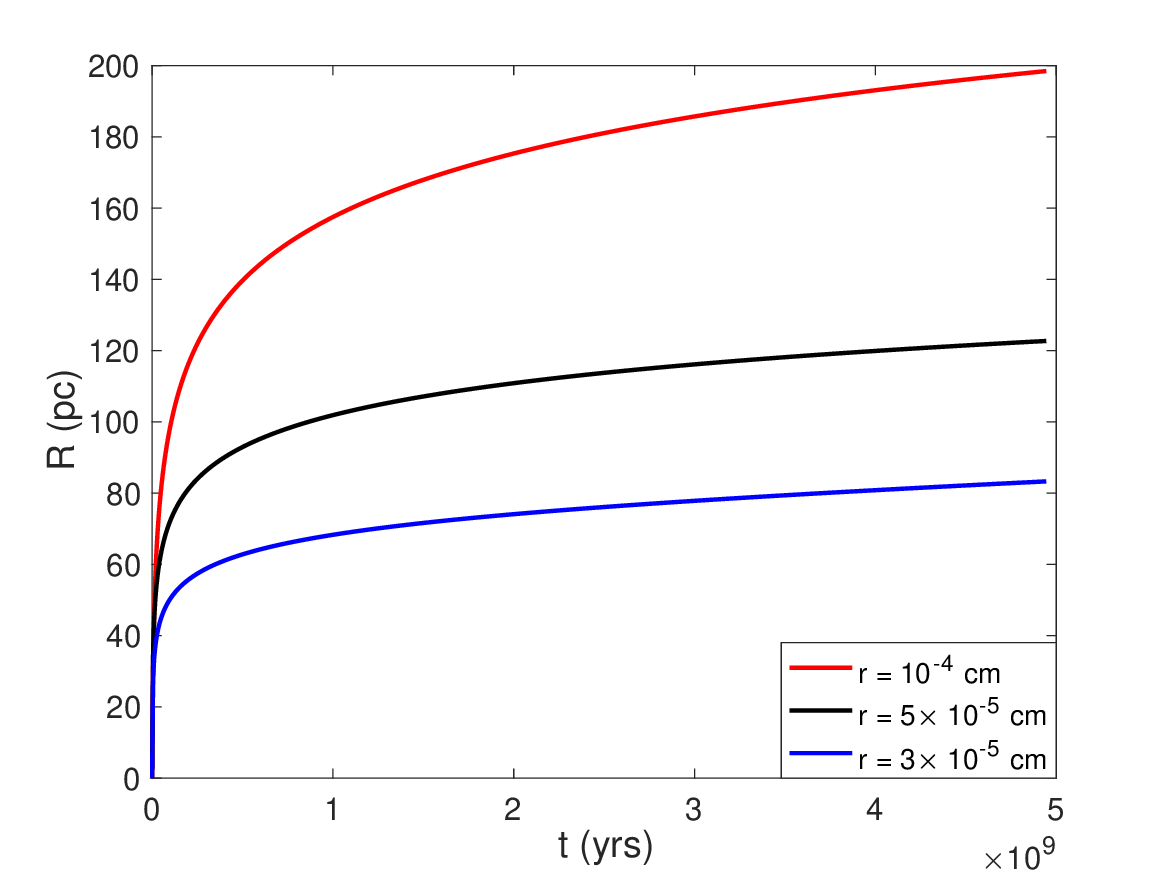}}
  \caption{Dependence of distance (in light years) of planetary dust particles versus time (in years) for three different values of radii: $r = 3\times10^{-5}$ cm (blue), $r = 5\times10^{-5}$ cm (black), and $r = 10^{-4}$ cm (red). The other set of parameters is: $M\simeq 2\times 10^{33}$ g, $L \simeq 3.83\times 10^{33}$ erg/s, $D = 1$, $\rho = 2$ g/cm$^3$, $n_0 = 1$ cm$^{-3}$ $R(0) \simeq 1$ AU, $\dot{R}(0)\simeq 0$ km/s, and $\upsilon_{\phi}\simeq 30$ km/s.}\label{fig1}
\end{figure}

After presenting the acceleration components in polar coordinates $a_{\phi} = R\ddot{\phi}-2\dot{R}\dot{\phi}$, $a_R = \ddot{R}-R\dot{\phi}^2$ \cite{carroll} and taking the geometry into account (see Fig. 1), one can straightforwardly obtain the equations of motion
\begin{equation}
\label{aphi} 
\ddot{\phi}\simeq -2\dot{\phi}\frac{\dot{R}}{R}-\frac{3D\rho_0}{4\rho r}\dot{\phi}\left(\dot{R}^2+R^2\dot{\phi}^2\right)^{1/2},
\end{equation}
$$\ddot{R}\simeq R\dot{\phi}^2-\frac{3D\rho_0}{4\rho r}\dot{R}\left(\dot{R}^2+R^2\dot{\phi}^2\right)^{1/2}+$$
\begin{equation}
\label{aR} 
+
\frac{1}{R^2}\left(\frac{3L}{16\pi c\rho r}-GM\right).
\end{equation}
We have assumed $m = 4\pi r^3\rho/3$, where $\rho$ denotes the mass density of the dust particle.

Motivated by panspermia, we focus on collisions between cosmic dust and planetary dust particles that may impart sufficient velocity to the latter to overcome Earth’s gravitational pull. \cite{berera}. It is important to emphasize that particles moving toward Earth will inevitably fall onto the planet—a purely geometric consideration—while, on average, half of those with sufficient velocity will escape into space.

However, our focus is on particles that do not experience significant heating, as this could damage any associated living organisms or complex molecules such as DNA or RNA. For the corresponding maximum velocity, $\upsilon_{m}$, the drag force power should be of the order of the thermal emission power of the heated dust particle $D\rho_a\pi r^2\upsilon^3\simeq 4\pi r^2\sigma T^4$ leading to 
\begin{equation}
\label{power} 
v_m\simeq\left(\frac{4\sigma T^4}{\rho_a D}\right)^{1/3},
\end{equation}
\newline
where $\sigma$ denotes the Stefan-Boltzmann's constant, $T\simeq 300$ K is the temperature when the complex molecules can survive, and $\rho_a$ represents the atmospheric mass density. For simplicity, we consider Earth parameters $\rho_a\simeq (1.2\times 10^{-3}\; g/cm^3)exp(-H/7.04)$ \cite{brekke,havens,picone}, where $H$ is the altitude measured in km. In \cite{berera} the author has noted that dust particles originating from an altitude of 150 km can attain velocities sufficient to escape Earth's gravitational field. From Eq. (\ref{power}) one can show that the maximum velocity relative to Earth, $v_m$, is on the order of 12.1 km/s, which is comparable to Earth's escape velocity, $11.2$ km/s. Therefore, for the initial conditions of the escaped planetary dust particles, one writes: $R(0) \simeq 1$ AU, $\dot{R}(0)\simeq 0$ km/s and $\upsilon_{\phi}\simeq 30$ km/s. Fig. 2 illustrates the time evolution of the radial distance of planetary dust grains for three different particle sizes. The set of parameters is: $M\simeq 2\times 10^{33}$ g, $L \simeq 3.83\times 10^{33}$ erg/s, $D = 1$, $\rho = 2$ g/cm$^3$, $n_0 = 1$ cm$^{-3}$ $R(0) \simeq 1$ AU, $\dot{R}(0)\simeq 0$ km/s, and $\upsilon_{\phi}\simeq 30$ km/s. 

In this study, we consider solar-type stars. As shown in the plots, over a period of $5$ billion years, planetary dust particles could travel considerable distances, $R_{max}$, of the order of $80$ pc ($r = 3\times 10^{-5}$ cm), $120$ pc ($r = 5\times 10^{-5}$ cm) and $200$ pc ($r = 10^{-4}$ cm). It is straightforward to show that the effects of the Brownian motion are negligible compared to the obtained results. The corresponding distance perpendicular to the total velocity might be approximated if one takes the mean free path of the dust grains $\lambda\simeq 1/(\pi r^2n_0)$ and the resulting time-scale between scatterings, $\tau\simeq\lambda/v_{rms}$. Here $v_{rms,_{\perp}}\simeq\sqrt{2kT/m}$ is the approximate thermal velocity in the perpendicular direction. Therefore, a one-step length scale is given by $d\simeq\tau v_{rms,_{\perp}}$, leading to the distance after S-steps: $D_{\perp}\simeq d\sqrt{S}$ \cite{carroll}. By substituting the physical parameters that we used for Fig. 1, one can show that $D_{\perp}$ is smaller than $R$ by many orders of magnitude, implying that the aforementioned distances are very realistic. Small irregularities may result in the rotation of dust particles, although the maximum rotational velocity is likely to be on the order of $\upsilon_m$, which will not change the overall temperature distribution of the dust particle.

If one takes into account the number density of G-type stars (solar type stars) in the solar neighborhood, $n_{st}\simeq 3.2\times 10^{-3}$pc$^{-3}$ \cite{bovy}, one can show that the planetary dust grains have reached $N_{st}\simeq 4\pi R_{max}^3n_{st}/3$ stars. In particular, $7\times 10^3$ stars for dust grains with $r = 3\times 10^{-5}$ cm, $2.5\times 10^4$ stars for $r = 5\times 10^{-5}$ cm, and $10^{5}$ stars for $r = 10^{-4}$ cm. 

Within the framework of this paper, we assume that the majority of the space in the ISM with temperatures ranging from $50$ K to $100$ K is primarily composed of molecular hydrogen. with a number density of the order of $n_0\simeq 1$ cm$^{-3}$ \cite{shkl,carroll}. These parameters can be favorable for preventing significant damage to complex molecules (DNA, RNA, etc.).

\section{UV damage}

A key aspect of this study is to understand the damage caused by UV radiation with wavelengths of $200-300$ nm. First of all, we need to find out what dose of radiation naked living organisms will receive while moving in space. For this purpose, we first consider the movement of a dust particle from the vicinity of a star to a distance, $R^*$, where the intensity of solar radiation 
\begin{equation}
\label{solar} 
I_S = \frac{\pi R_{\odot}^2}{R^{*2}}\int_{UV}B_{\lambda}d\lambda,
\end{equation}
and the total background intensity from all stars in distant space
\begin{equation}
\label{total} 
I_b = 4\pi W\int_{UV}B_{\lambda}d\lambda,
\end{equation}
become of the same order. Here 
\begin{equation}
\label{BB} 
B_{\lambda} = \frac{2 hc^2}{\lambda^5\left(\exp\left(hc/(kT\lambda)\right)-1\right)}
\end{equation}
is the Planck's black body radiation function, $\lambda$ denotes the wave-length of the emission spectrum, $h$ is the Planck's constant, $R_{\odot}\simeq 7\times 10^{10}$ cm denotes the solar radius and $W = 10^{-14}$ \cite{scheffler}. From the above equations, one can readily determine that $R^* = 3.5\times 10^{17}$ cm. The UV radiation dose is defined as \cite{wesson}
\begin{equation}
\label{dose} 
D =\pi R_{\odot}^2 \int_0^{\tau} \frac{dt}{R(t)^2} \int_{\lambda_1}^{\lambda_2}B_{\lambda}d\lambda,
\end{equation}
where $\tau$ is the time-scale associated with $R^*$ and $\lambda_1 = 1600$ A, $\lambda_1 = 3000$ A. Taking into account the solutions of Eqs. (\ref{aphi},\ref{aR}) an estimate of the dose value can be obtained. $D\simeq 6\times 10^{15}$ erg/cm$^2$. This value is so high that no naked bacteria can survive the trip. A living organism encased in a protective mantle would be shielded from the star's radiation. However, despite the extremely low background ultraviolet radiation, the travel time through space could be extensive, leading to a potential increase in the radiation dose, which may reach critical levels. For studying this particular task, one should introduce the inactivation cross-section, $\tilde\sigma$ defining the fraction of the surviving irradiation  
\begin{equation}
\label{frac} 
\frac{N}{N_0} = e^{-\tilde\sigma D}.
\end{equation}
It is evident that the inactivation of a bacterium is characterized by the condition  $\tilde\sigma D\simeq 1$.

However, the inactivation cross-section varies for different wavelengths. In particular, for the T1 bacteriophage $\tilde\sigma_{_1}^*=10^{-15.07058}$ cm$^2$/photon for $1600 A<\lambda_A<2262A$, $\tilde\sigma_{_2}^*=10^{-2.67-0.005481\lambda_A}$ cm$^2$/photon for $2262 A<\lambda_A<2350 A$,  $\tilde\sigma_{_3}^*=10^{-86.05+0.05578\lambda_A-1.061\times 10^{-5}\lambda_A^2}$ cm$^2$/photon for $2350 A<\lambda_A<3000 A$, where by $\lambda_A$ we denote the wavelength in units of angstroms \cite{wesson}. Then, after taking into account a natural conversion $\tilde\sigma = \lambda\tilde\sigma^*/hc$, one can write a condition
$$(4\tilde\tau W)^{-1}\simeq $$
\begin{equation}
\label{inac} 
\simeq\int_{\lambda_1}^{\lambda_2}\tilde\sigma_1\tilde B_{\lambda}d\lambda+\int_{\lambda_2}^{\lambda_3}\tilde\sigma_1\tilde B_{\lambda}d\lambda+\int_{\lambda_3}^{\lambda_4}\tilde\sigma_1\tilde B_{\lambda}d\lambda,
\end{equation}
where $\lambda_1= 1600$ A, $\lambda_2= 2262$ A, $\lambda_3= 2350$ A, $\lambda_4= 3000$ A, $\tilde\tau$ is the timescale corresponding to the inactivation of a bacteria, $\alpha = exp\left(-4\pi\kappa r/\lambda\right)$ is a transmittance coefficient, $\kappa\simeq 1.7$ \cite{wesson} and $\tilde B_{\lambda} = \alpha B_{\lambda}$. 

\section{Discussions}
After taking the aforementioned radii of the grains, one can show that the inactivation time scales are: $\tau\simeq 1.2$ yr for $r = 3\times 10^{-5}$ cm, $\tau\simeq 200$ yrs for $r = 5\times 10^{-5}$ cm, and $\tau\simeq 10^7$ yrs for $r = 10^{-4}$ cm respectively. This means that living organisms might be transported on distances respectively $13AU$ for $r = 3\times 10^{-5}$ cm, $0.42$ pc for $r = 5\times 10^{-5}$ cm and $43$ pc for $r = 10^{-4}$ cm. Therefore, in the case of Earth, it is quite possible that life has been transported to other planets in the solar system. But the most fascinating scenario concerns the dust grains with $r = 10^{-4}$ cm, where the mantle can survive the living bacteria for $10^7$ yrs, implying that the grains might reach $10^3$ stars. It is worth noting that after inactivation, the complex molecules might still be transported to distant worlds and, as has already been emphasized, almost $10^5$ stars could be influenced by these grains.

Another issue we would like to address is the flux of dust particles transporting complex molecules to another planet. In \cite{berera}, the author analyzing the data of space dust \cite{flux1,flux2,flux3,flux4,flux5} came to the conclusion that the flux of dust particles leaving the Earth's gravitational field is $\mathcal{F}_0\simeq 1$ cm$^{-2}$s$^{-1}$. Then, the flux on a distance, $R$, writes as $\mathcal{F}\simeq\mathcal{F}_0R_{\oplus}^2/R^2$, where $R_{\oplus}\simeq 6400$km is the Earth's radius. On the other hand, by taking a uniform distribution of stars into account, one can estimate the number of planets seeding a particular planet by dust particles, $N_l^{*}\simeq N_lN_{st}/N_*$, leading to the total rate on a planet's surface, $\mathcal{K}_{tot}\simeq 2\pi N_l^{*}\mathcal{F_0}R_{\oplus}^4/R^2\simeq 3\times 10^{-8}$s$^{-1}$. This implies that, each year, one particle could reach the planet, resulting in billions of dust grains accumulating over a billion years for each individual planet.

It is worth noting that relatively dense regions of the Milky Way, known as molecular clouds, may serve as natural obstacles for dust particles, which might be trapped by them. These clouds are known to be regions composed mostly of hydrogen atoms or molecules \cite{carroll}. In particular, it is well known that in diffuse H I clouds, with temperatures $(30, 80)$ K the number density varies in the interval $(1-8)\times 10^2 cm^{-3}$. In diffuse molecular clouds, typical temperatures lie in the range $(15, 50)$ K with $n\simeq (0.5-5)\times 10^3 cm^{-3}$. Giant molecular clouds exhibit a temperature of the order of $15$ K and $n\simeq (1-3)\times 10^2 cm^{-3}$. Dark-cloud complexes have more or less the same values of $T$ and $n$, but small individual clumps with $T\simeq 10$ K might be even denser $10^{3-4} cm^{-3}$. In hot cores $T\simeq (100-300) K$ the number density could be significantly higher compared to the previous cases $10^{9-11} cm^{-3}$ \cite{carroll}.

If dust grains encounter these regions, one can estimate the distance $S_d$, which the particles will travel within the clouds. Leaving only the drag force in Eq. \ref{aR} one can easily show $S_d\simeq \frac{4r\rho}{3D\rho_c}\ln(v_0/v_{rms})$, where $v_0$ is the initial velocity of the dust particles entering the clouds, $\rho_c\simeq n_cm_c$, $m_c=m_p, 2m_p$ and we assume that when the velocity of the dust particles increases on the order of $v_{rms}\simeq\sqrt{3kT/m_c}$, the particles are trapped by the clouds. Considering the hot cores having the smallest characteristic sizes $0.05$ pc \cite{carroll}, even for $v_0\simeq 30$ km/s, the distance becomes by several orders of magnitude less than the aforementioned length. This is a very important result because it indicates that dense molecular clouds might contain planetary dust grains with complex molecules inside. But the whole trip of the dust grains is not that simple. It is not enough to consider only UV radiation to
analyze the possible transport of life through the galaxy. If the dust particles encounter an X-ray radiation region, the molecules will be completely destroyed. 

On the other hand, extremely hot regions, such as ionized HII regions in stellar atmospheres and high-temperature areas in the ISM with temperatures ranging from $10^6$ K to $10^7$ K will inevitably destroy molecules. Additionally, it remains unclear how bacteria would behave under such extreme conditions, especially if these conditions persist for hundreds of years or more, therefore, further research is essential.

\section{Summary}
We have considered the dynamics of planetary dust particles "propelled" by planets, and it has been shown that in $5$ billion years, dust grains can travel in the ISM at distances of the order of hundreds of parsecs.

By taking the stellar distribution density into account, we have found that dust particles emitted by every single planet will reach as much as $10^5$ stellar systems.

It has been found that life can be transported to $\sim 1000$ stars only from a single planet.

We have also shown that dense molecular clouds will efficiently trap the planetary dust grains and their study in the context of panspermia might be significant.

Several key factors that contribute to the damage of life and complex molecules associated with living organisms have been analyzed. In particular, we have emphasized that, in addition to UV radiation, the X-ray background even more critically can inevitably inactivate bacteria and damage complex molecules. 

Another important factor is the behavior of bacteria in vacuum conditions over hundreds of years or more - a topic that remains poorly understood and requires further investigation.

\section*{Acknowledgments}
This work was supported by the Shota Rustaveli National Science Foundation of Georgia under Grant No. FR-24-1751. The publication charges have been supported by the Knowledge Foundation. I would like to thank I. Sonishvili for their invaluable support, insightful discussions, and memorable collaboration.

\nolinenumbers





\end{document}